\documentclass[%
superscriptaddress,
amsmath,amssymb,
aps,
prb,
twocolumn,
byrevtex,
floatfix
]{revtex4-2}

\usepackage{graphicx}
\usepackage{dcolumn}
\usepackage{bm}
\usepackage{multirow}
\usepackage{float}
\usepackage{overpic}
\usepackage{subcaption}
\usepackage{caption}
\usepackage{makecell}
\usepackage{oplotsymbl}
\usepackage{todonotes}
\usepackage{hyperref}
\usepackage{color}
\usepackage{graphpap}
\usepackage{MnSymbol}
\usepackage{ulem}

\begin{document}

\preprint{APS/123-QED}

\title{Influence of interstitial Li on the electronic properties of Li$_{x}$CsPbI$_{3}$ for \\photovoltaic and battery applications}

\author{Wei Wei}
\affiliation{Freiburg Center for Interactive Materials and Bioinspired Technologies, University of  Freiburg, Georges-K\"ohler-Allee 105, 79110 Freiburg, Germany}
\affiliation{Fraunhofer Institute for Mechanics of Materials IWM, W\"ohlerstra{\ss}e 11, 79108 Freiburg, Germany}

\author{Julian Gebhardt}
\affiliation{Fraunhofer Institute for Mechanics of Materials IWM, W\"ohlerstra{\ss}e 11, 79108 Freiburg, Germany}

\author{Daniel F. Urban}
\affiliation{Fraunhofer Institute for Mechanics of Materials IWM, W\"ohlerstra{\ss}e 11, 79108 Freiburg, Germany}
\affiliation{Freiburg Materials Research Center, University of Freiburg, Stefan-Meier-Stra{\ss}e 21, 79104 Freiburg, Germany}

\author{Christian Elsässer}
\email{christian.elsaesser@iwm.fraunhofer.de}
\affiliation{Freiburg Center for Interactive Materials and Bioinspired Technologies, University of  Freiburg, Georges-K\"ohler-Allee 105, 79110 Freiburg, Germany}
\affiliation{Fraunhofer Institute for Mechanics of Materials IWM, W\"ohlerstra{\ss}e 11, 79108 Freiburg, Germany}
\affiliation{Freiburg Materials Research Center, University of Freiburg, Stefan-Meier-Stra{\ss}e 21, 79104 Freiburg, Germany}

\date{\today}

\begin{abstract}
The integrated device of a perovskite solar cell with a Li-ion battery is an innovative solution for decentralized energy storage in smart electronic devices. In this study, we examine the stability of Li ions intercalated in a CsPbI$_3$ perovskite and their effect on the electronic structure of Li$_x$CsPbI$_3$ compounds using first-principles density functional theory. Our simulations demonstrate that the insertion of Li at concentrations up to $x$ = 1 into CsPbI$_3$ perovskite is energetically possible. Moreover, we identify that the distortion of the Pb-I octahedra has the strongest impact on the change in the electronic band gap. Specifically, an increase in the amount of intercalated Li causes larger structural distortions, which in turn lead to an increasing band gap as function of the Li content. 
\end{abstract}

\maketitle


\section{Introduction\protect}

Advances in the design of smart electronic devices pose significant challenges for the integration and miniaturization of solar cells and energy storage components  \cite{lai2017comprehensive,arora2018crafting,berestok2022monolithic,zhang2020halide}. A promising approach involves the incorporation of halide perovskites as photovoltaic absorbers and Li-ion battery electrodes in commercial applications. Halide perovskites possess mixed ionic-covalent bonding character  \cite{li2017extrinsic,zhang2018extrinsic}, unlike the covalent or polar semiconductor compounds that are typically used in photovoltaic devices, and show good ionic conductivity \cite{li2017extrinsic}.
Therefore, halide perovskite materials, which were originally designed for solar cells, may serve as multifunctional materials for both, light harvesting and Li storage. 

The maximum amount of Li ions that can be stored in the unit cell of a perovskite crystal is determined by the quantity of available interstitial sites for the Li ions and their respective binding energies. Understanding and improving the performance of perovskite-based electrode materials for Li-ion batteries requires the knowledge of how the stability of materials is influenced by the intercalation of Li. Despite its importance, the interaction between inserted Li ions and the perovskite host crystal structure has not been studied extensively so far. Several experimental studies have looked at the structure and stability of Li-perovskite systems, and have reported a wide range of Li uptake limits of $x=0$ -- 6 in Li$_x$CsPbI$_3$  \cite{buttner2022halide,xia2015hydrothermal,vicente2017methylammonium,dawson2017mechanisms}. However, this wide range has not yet been comprehensively interpreted in the literature.

Furthermore, the insertion of interstitial elements always alters the electronic structure of the host crystal \cite{willoughby1978atomic} which will have an influence on the performance of the solar cell \cite{henry1980limiting}. While band gaps of Li-perovskite material systems have been measured experimentally  \cite{jiang2017electrochemical,mathieson2022solid}, the relationship between the Li concentration and the band gap has not yet been systematically studied or explained.

In this work, we present a computational study based on density functional theory (DFT) to investigate the effects of Li-ion intercalation in Li$_x$CsPbI$_3$ as a prototype. CsPbI$_3$ is a well studied reference compound for inorganic halide-perovskite materials, with good light-harvesting capabilities  \cite{mahato2020highly,zhang2019improved,kye2019vacancy,hoffman2016transformation,sutton2018cubic,wang2019energetics}. We investigate two structural scenarios that serve as limiting cases for the dynamic CsPbI$_3$ perovskite structure under device-operation conditions \cite{wei2024location}. The two cases, namely the cubic $\alpha$ structure and the distorted $\gamma$' structure, represent the highest saddle point configuration and the configurations at the local energy minima (when tetragonal lattice distorsions are neglected) of the vibrating CsPbI$_3$ crystal, respectively. We study the Li uptake limit in the CsPbI$_3$ perovskite structure for both cases, and the influences of Li on the crystal and electronic structures of CsPbI$_3$ are systematically explored.

The paper is organized as follows. Sec. \ref{computational} introduces the computational setup and the definition of the formation energy of Li inserted into CsPbI$_3$. In Sec. \ref{results}, we present the results on the stability of interstitial Li within CsPbI$_3$ and its influence on the electronic structure. In Sec. \ref{discussion}, we discuss our results in view of other findings in the literature, and we summarize and conclude in Sec. \ref{summary}.

\section{Computational Details} \label{computational}
 
All calculations are carried out with the Vienna ab initio simulation package (VASP) \cite{kresse1996efficiency} employing projector-augmented waves \cite{blochl1994projector} and the strongly constrained and appropriately normed (SCAN) meta-generalized gradient approximation functional \cite{sun2015strongly,sun2016accurate}. An energy cutoff of 520 eV was used for the plane-waves basis. Total-energy differences and forces on atoms for all structural degrees of freedom are converged within $1 \times10^{-5}$ eV and $5 \times10^{-3}$ eV/\AA, respectively. The Brillouin-zone integrals were sampled by 4$\times$4$\times$4 Monkhorst-Pack $k$-point grids \cite{monkhorst1976special} with a Gaussian smearing of $1\times10^{-3}$ eV for
the 2$\times$2$\times$2 supercell models, containing 40 atoms (i.e., eight ABX$_3$ formula units). Structural relaxations for these systems are carried out in internal coordinates \cite{buvcko2005geometry}. The implementation was improved to retain the initially set symmetry of the crystal. 

To obtain accurate crystal structures, we combined the SCAN functional with the nonlocal correlation functional rVV10 \cite{sabatini2013nonlocal} to account for van der Waals (vdW) dispersion interactions in our DFT calculations for the perovskite structures. Numerous authors have reported that the inclusion of vdW interactions in DFT calculations is useful to get quantitatively more accurate structural parameters for crystals, as compared to experimental data; see, e.g. the work of Xue et al. \cite{xue2021first,xue2023compound} or other previous studies \cite{jing2019tuning,fadla2020first,gebhardt2021efficient}. In our present work, the obtained values for the equilibrium lattice constant of CsPbI$_3$ are $a_0$(PBE)=6.39~\AA~\cite{wei2024location} and $a_0$(SCAN+rVV10)=6.27~\AA. The SCAN+rVV10 result agrees better than the PBE result with reported experimental data that vary from $a_0$(Expt.)=6.18~\AA~to 6.30~\AA~\cite{eperon2015inorganic,trots2008high,marronnier2018anharmonicity}. 

The defect formation energies of Li ions intercalated at interstitial sites in the perovskite crystal are calculated as 
\begin{equation}
	E_{\rm form}=\frac{E_{\rm tot}[{\rm Li_xCsPbI_3}] - E_{\rm tot}[{\rm CsPbI_3}] - x_{\rm Li} E_{\rm tot}[{\rm Li}_{\rm bcc}]}{x_{\rm Li}} \,,
\end{equation}
where $E_{\rm tot}[{\rm Li_xCsPbI_3}]$ is the total energy of the considered perovskite crystal containing Li, $E_{\rm tot}[{\rm CsPbI_3}] $ is the energy of the reference crystal without Li, $E_{\rm tot}[{\rm Li}_{\rm bcc}]$ is the energy per Li atom in the body-centered cubic one-atom unit cell of the elemental Li metal, and $x \rm _{Li}$ is the proportion of Li per formula unit of the perovskite crystal. We assume that the perovskite crystal is in a Li-rich environment, i.e. in contact with a thermodynamic reservoir of metallic (body-centered cubic, BCC) Lithium. Therefore, the reference chemical potential is the energy per atom of the bulk BCC Li metal. The calculated lattice constant of BCC Li is 3.42~\AA, and the total energy per atom of this metal was calculated using a 10$\times$10$\times$10 Monkhorst-Pack k-point grid and otherwise the same computational parameters as for the perovskite. 

For calculating the electronic structure, we employ the DFT+1/2 method \cite{ferreira2008approximation} including spin-orbit coupling \cite{steiner2016calculation}. This method provides a similarly accurate electronic-structure description as hybrid-functional methods for CsPbI$_3$ perovskites, as demonstrated in our previous study \cite{gebhardt2021efficient}.

\section{Results} \label{results}

\subsection{The stability of interstitial Li in CsPbI$_3$} \label{A}

\begin{figure}
	\centering
	\begin{overpic}[width=0.50\textwidth]{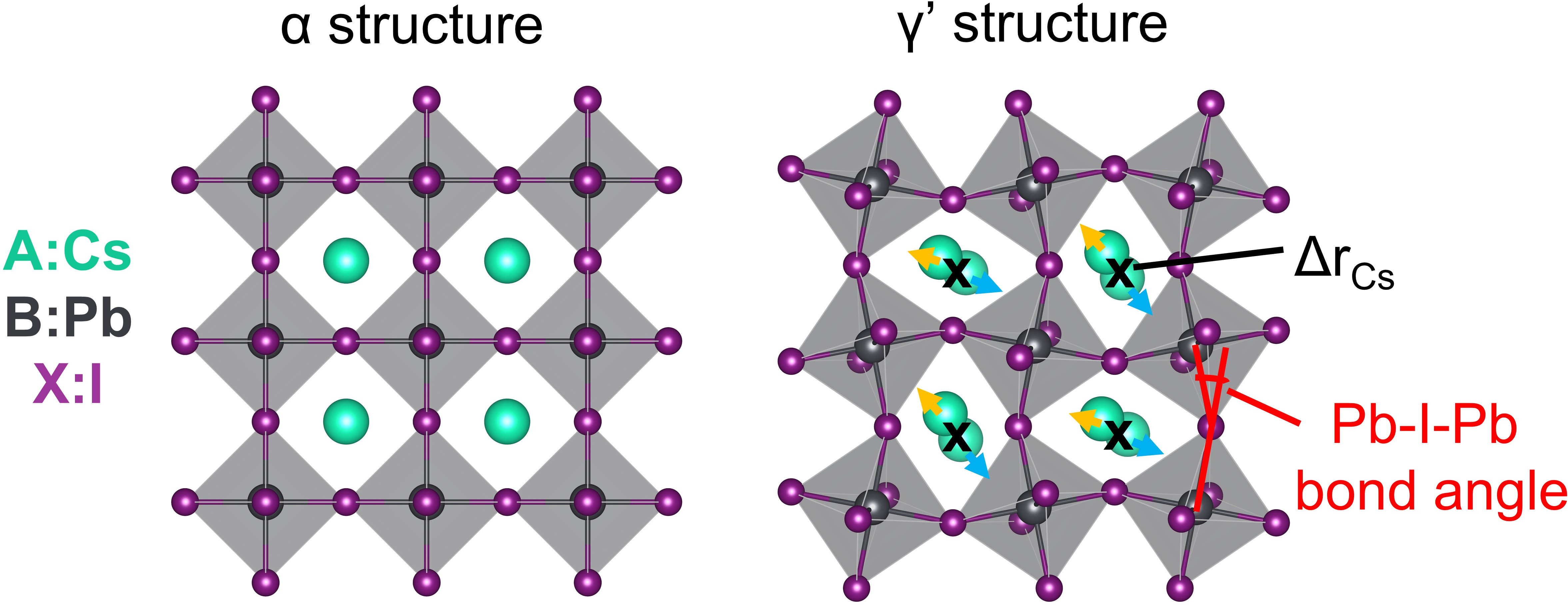}
	\end{overpic}
	\caption{The two limiting models of the dynamic	crystal structure of CsPbI$_3$, illustrating the degrees of structural distortion. The B-site atoms are located on the lattice points of the cubic (pseudo-cubic) structure. The off-center displacement of the A-site atoms ($\Delta {\rm r_{Cs}}$) and the tilts of B-X octahedra described by the averaged Pb-I-Pb bond angle difference ($\Delta_{\text{Pb}\mbox{-}\text{I}\mbox{-}\text{Pb}}$) quantify the structural distortion. 
	}
	\label{fig1}
\end{figure}

We studied two structural models, the $\alpha$ and $\gamma$' structures, to represent the two limiting cases of the dynamical CsPbI$_3$ phase at finite temperature. Figure \ref{fig1} illustrates these two models. The ordered cubic $\alpha$ structure is observed as a temporal and spatial average in experimental studies \cite{whitfield2016structures,yang2017spontaneous} and represents the highest saddle-point energy configuration of the phase, while the disordered $\gamma$' structure is obtained as the configuration of the local energy minimum of the $\alpha$ phase. 
Note that in reality the tetragonal lattice distorsions towards the $\gamma$ phase, which are observed at lower temperatures, are likely to occur in an averaged manner at higher temperatures, too. However, the influence of the concomitant small structural changes is weak, \cite{wei2024location,klarbring2019low} (e.g., the $\gamma$ phase is more stable than the $\gamma$’ structure by only 4.2 meV per unit cell) and, therefore, neglected for the remainder of this study.
The structural difference between $\alpha$ and $\gamma$' can be attributed to two factors: the off-center displacement of the A-site Cs atoms ($\Delta {\rm r_{Cs}}$), and the octahedral B-X tilts that occur around two axes \cite{marronnier2018anharmonicity} in the ABX$_3$ compound CsPbI$_3$. Here, we quantify these octahedral tilts by the averaged Pb-I-Pb bond angle difference ($\Delta_{\text{Pb}\mbox{-}\text{I}\mbox{-}\text{Pb}}$).

In the following, we study the stability of interstitial Li as function of its concentration in the two structural models using a 2$\times$2$\times$2 supercell. By subsequently adding Li to the two models, Li$_x$CsPbI$_3$ was obtained with Li concentrations varying from $x$=1/8 to $x$=1 for the $\alpha$ structure, and $x$=1/8 to $x$=2 for the $\gamma$' structure. We analyzed various different Li arrangements in the two models, as Li-Li interactions need to be considered for Li concentrations above the dilute limit.

\subsubsection{The stability of interstitial Li in the $\alpha$ structure}

\begin{figure}
	\centering
	\begin{overpic}[width=0.30\textwidth]{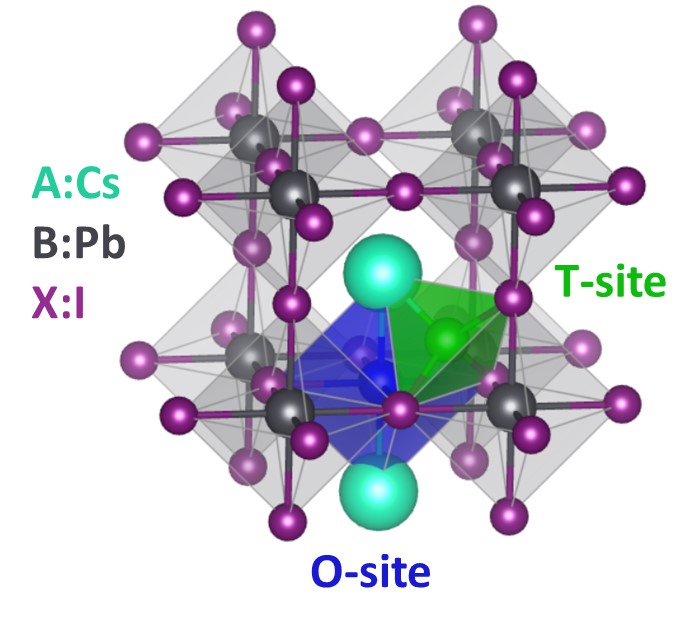}
	\put(-10,85){(a)}
    \end{overpic}
	\begin{overpic}[width=0.45\textwidth]{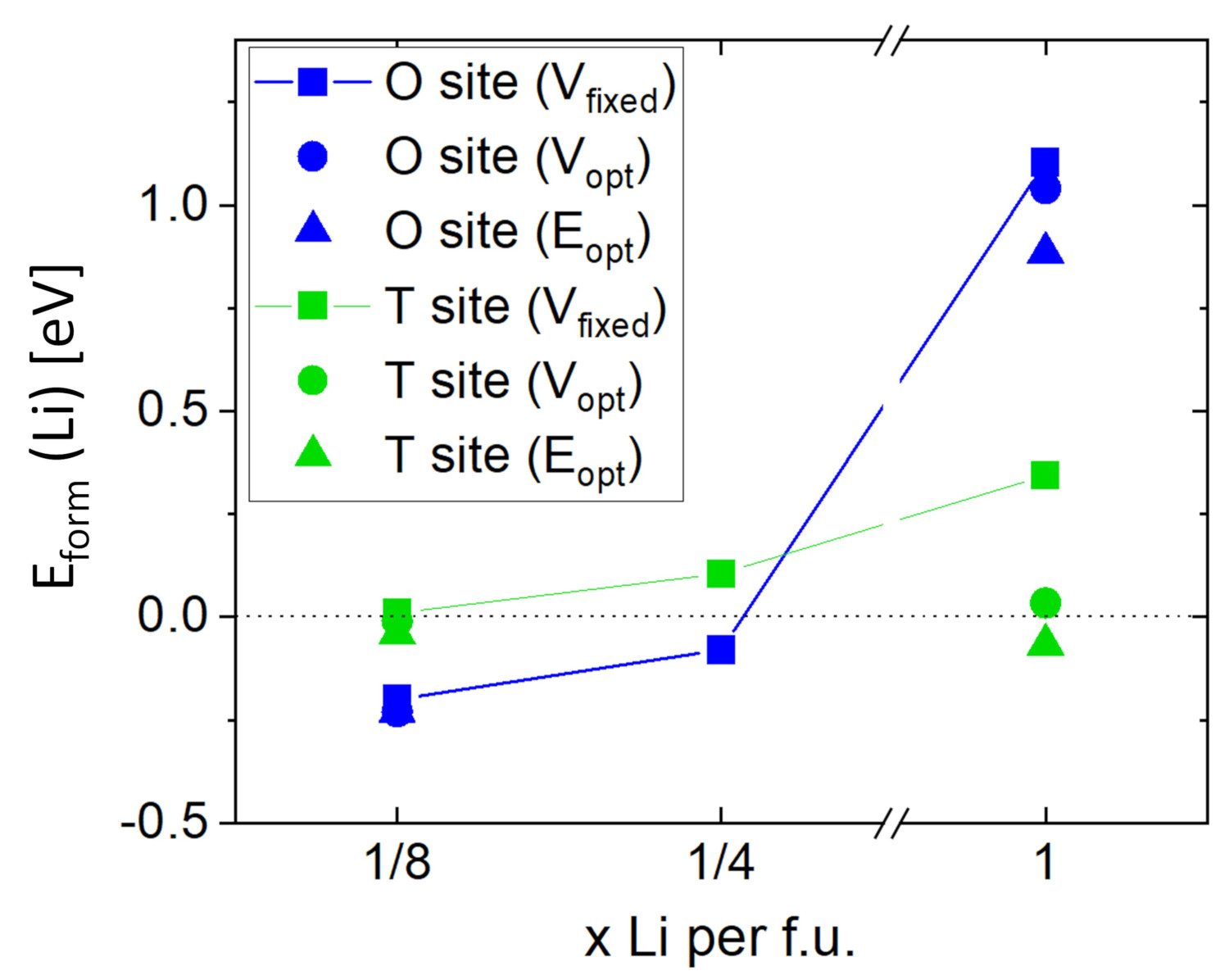}
	\put(5,75){(b)}
    \end{overpic}
	\caption{a) Li interstitial sites in the $\alpha$ structure, located in the center of an octahedron (O) or a tetrahedron (T) formed by host lattice atoms. b) Formation energies of interstitial Li as function of the concentration $x$. The horizontal dotted line labels 0 eV. Square symbols indicate results obtained for a fixed simulation cell ($V_\text{fixed}$), circle symbols indicate that the volume of the cell was optimized, but the rectangular shape of the cell was maintained during optimization (V$_\text{opt}$), and the triangle symbols indicate structural relaxation in all degrees of freedom (E$_\text{opt}$). The point symmetry of the interstitial Li at the O site ($D_{4h}$) or T site ($C_{3v}$) is maintained for all Li concentrations. }
	\label{fig2}
\end{figure}

In the $\alpha$ structure, the high-symmetry interstitial sites are three octahedral (O) and eight tetrahedral (T) sites as displayed in Figure~\ref{fig2}a. The O site is located in the center of an octahedron formed by two Cs atoms and four I atoms, while the T site is located in the center of a tetrahedron formed by one Cs atom and three I atoms.
 
Figure~\ref{fig2}b shows the formation energies of Li in both interstitial sites ranging from {$-0.20$} to 1.10 eV for $x$=1/8 to $x$=1. We study the influence of varying the Li concentration $x$ by: i) excluding changes to the lattice constant ($V_\text{fixed}$), ii) optimizing the volume (V$_\text{opt}$), and iii) allowing full relaxation of unit-cell degrees of freedom (E$_\text{opt}$) for each $x$. The highest point symmetry of a Li atom on an O site ($D_{4h}$) or a T site ($C_{3v}$) is maintained at every concentration $x$ in all three cases.

For the fixed cell at low concentrations of $x$=1/8 and $x$=1/4, the insertion of Li atoms at the O sites is energetically more favorable by approx.~0.2 eV compared to the T sites. At $x$=1/8, the E${\rm _{form}}$ on the O site is {$-0.20$} eV, and an optimization of lattice degrees of freedom does not further stabilize the interstitial site. Increasing the concentration to $x$=1/4 rises E${\rm _{form}}$ to {-0.08} eV for the O site. In contrast, the results for $x$=1 differ significantly from those for the two lower concentrations. The order of E${\rm _{form}}$ of the two sites is reversed and, compared to $x$=1/4, the formation energy of Li is increased by at least 0.42 eV (when changes in lattice parameters are excluded). 
	Additionally, variations in lattice volume and shape have an influence on the E${\rm _{form}}$ at x=1. Compared to keeping the volume fixed (V$_\text{fixed}$), allowing the volume to adjust to the Li content (V$_\text{opt}$) results in a stabilization of 0.06 eV for the O site and 0.31 eV for the T site (energy differences between square and circle symbols at $x=$1 in figure~\ref{fig2}b), accompanied by an increase of the volume of 9\% and 17\%, respectively. 
Structural distortions away from the cubic structure, by relaxing all lattice parameters (E$_\text{opt}$)
 are energetically stabilized further by 0.016~eV 
 	for the O site and 0.10~eV for the T site (energy differences between circle and triangle symbols at $x=$1 in figure~\ref{fig2}b). The lattice undergoes a tetragonal distortion with the insertion of Li at the O site, while the insertion of Li at the T site transforms the lattice into a rhombohedral cell with a cell-edge angle of 81$^\circ$.

\begin{figure}
	\centering
	\begin{overpic}[width=0.5\textwidth]{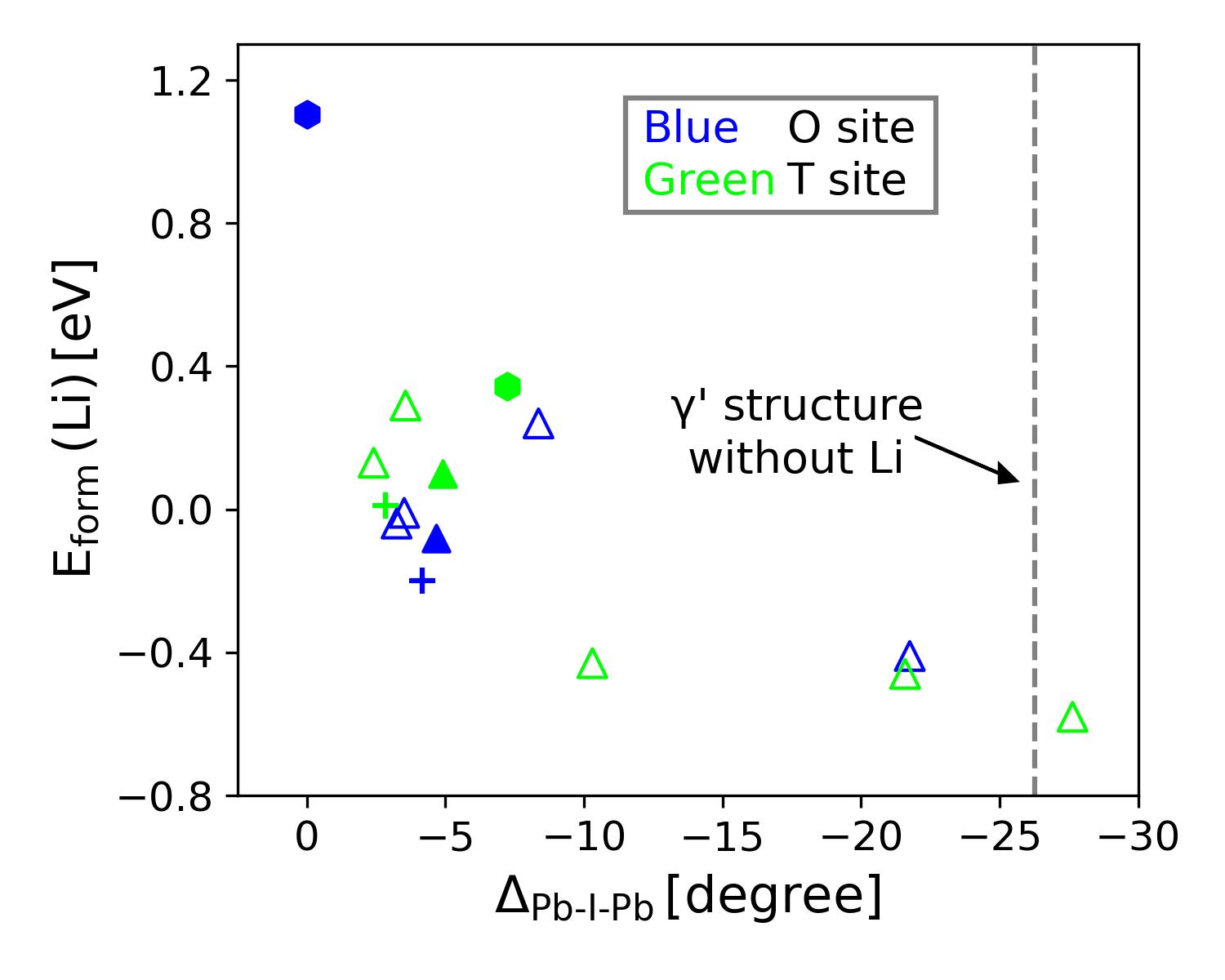}
	\end{overpic}
	\caption{The relationship between the formation energy of interstitial Li and the structural distortion $\Delta_{\text{Pb}\mbox{-}\text{I}\mbox{-}\text{Pb}}$ with varying Li concentration $x$ in the $\alpha$ structure at fixed volume and cell shape. Blue and green symbols indicate interstitial Li on O and T sites, respectively. 
		The different Li concentrations are indicated by these symbols: "+" for x=1/8, $\triangle$ for x=1/4, and $\hexago$ for x=1.
		Filled and open symbols represent high-symmetry and low-symmetry structures of the Li$_x$CsPbI$_3$ compounds, respectively. 
	}
	\label{fig3}	
\end{figure}

While the previous paragraphs apply to Li arrangements keeping the high symmetry of the host crystal structure, we now consider various Li arrangements at $x$=1/4 without enforcing any symmetry. The different arrangements of Li atoms influence their formation energy in the perovskite crystal. We observed that the formation energy is primarily associated with structural distortions of the host lattice, specifically characterized by a decrease of the formation energy with an increase of $\Delta_{\text{Pb}\mbox{-}\text{I}\mbox{-}\text{Pb}}$. This relationship even exceeds the changes concentration brings to the formation energy. As we observed in Figure \ref{fig3}, this relationship holds for all concentrations and Li arrangements investigated for the $\alpha$ structure. As the structural distortion increases, the $\alpha$ structure transforms to the $\gamma$' structure, accompanied by a decrease in the formation energy.

\begin{figure}
	\centering
	\begin{overpic}[width=0.5\textwidth]{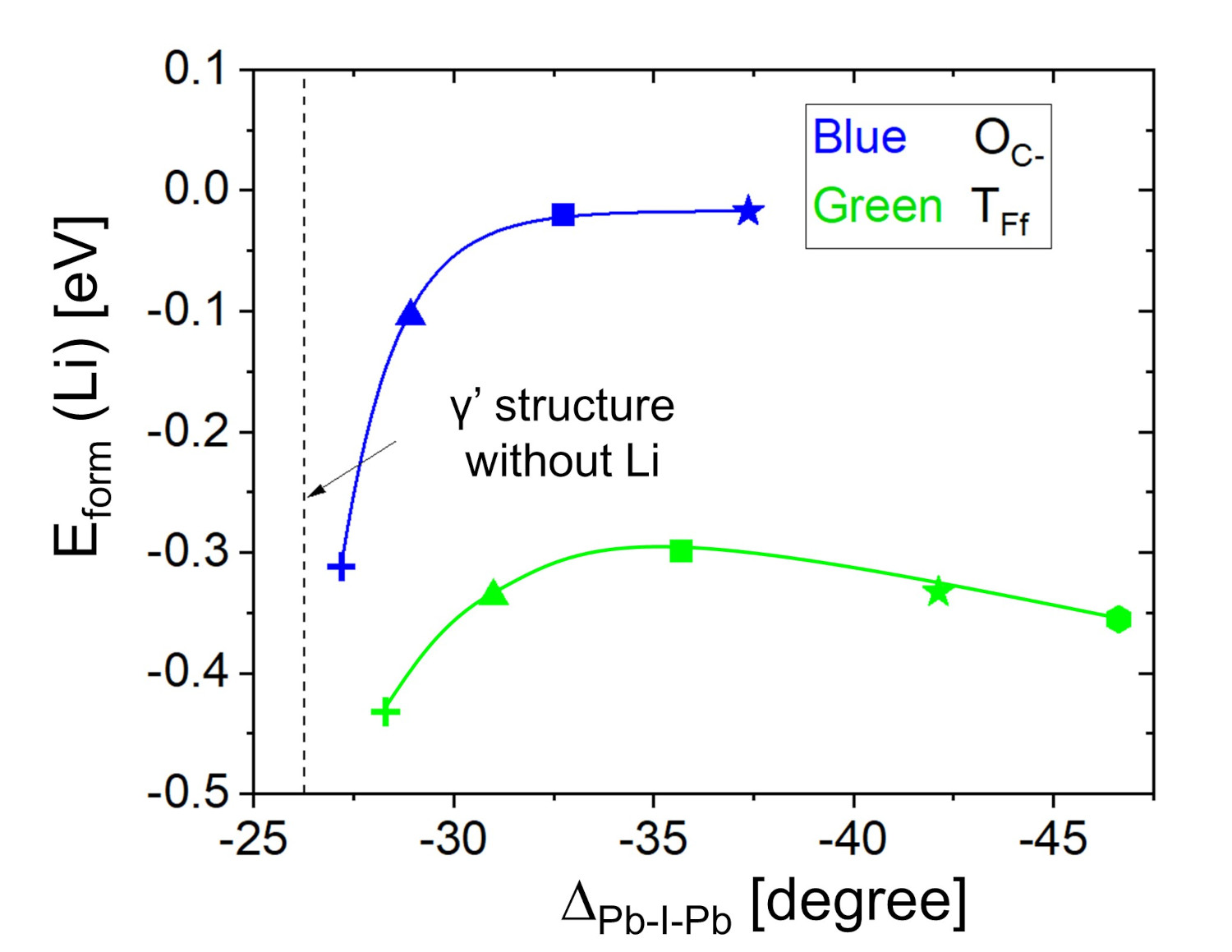}
	\end{overpic}
	\caption{The relationship between the formation energy of interstitial Li and the structural distortion $\Delta_{\text{Pb}\mbox{-}\text{I}\mbox{-}\text{Pb}}$ for the most stable O and T sites with varying Li concentration $x$ in the $\gamma$' structure. 
		The different Li concentrations are indicated by these symbols: "+" for x=1/8, $\triangle$ for x=1/4, $\square$ for x=1/2, $\largestar$ for x=3/4, and $\hexago$ for x=1. 
	}
	\label{fig4}	
\end{figure}

Therefore, the stability limit for interstitial Li in CsPbI$_3$ while maintaining the symmetry of the $\alpha$ structure is at $x$=1/4. For structures with $x$=1/2 it was impossible to maintain this high symmetry with any reasonable Li distribution.

\subsubsection{The stability of interstitial Li in the $\gamma$' structure}

Due to the lower symmetry of the $\gamma$’ structure, compared to the $\alpha$ structure, the six O and eight T interstitial sites are no longer symmetry equivalent. After optimization, the most stable T and O sites have formation energies of -0.43 and -0.31 eV per Li atom, respectively, in the 40-atoms supercell, i.e. at a Li concentration of x=1/8. We label these two most stable interstitial sites as T${\rm _{Ff}}$ and O${\rm _{C-}}$ in accordance with the nomenclature introduced in our preceding work \cite{wei2024location}.
 
Figure~\ref{fig4} summarizes the Li formation energy of interstitial Li in the $\gamma$’ structure for $x$= 1/8, 1/4, 1/2, 3/4, and 1, on the T${\rm _{Ff}}$ and O${\rm _{C-}}$ sites. In this model, the Li atoms are arranged as far away from each other as possible in the given supercell.  Higher concentrations of Li consistently result in stronger structural distortions, but do not always lead to higher formation energies. From $x$=1/8 to $x$=1/2, the formation energy increases with increasing structural distortion $\Delta_{\text{Pb}\mbox{-}\text{I}\mbox{-}\text{Pb}}$. When $x$ exceeds 1/2, $E_{\rm form}$ reaches a peak. By further increasing $x$ to 1 on T${\rm _{Ff}}$ sites, a slight decrease of E${\rm _{form}}$ is observed. However, Li is no longer stable on O${\rm _{C-}}$ sites in structures with $x$=1 and the Li atoms move to the neighboring T${\rm _{Ff}}$ sites during the geometry optimization. 
The energy profile is again consistent with the $\gamma$' structure being the energetically most favorable structure. The minimum energy is obtained for a Li content of $x=1/8$ for which the resulting crystal structure is closest to the $\gamma$' structure. 

\begin{figure}
	\includegraphics[width=0.5\textwidth]{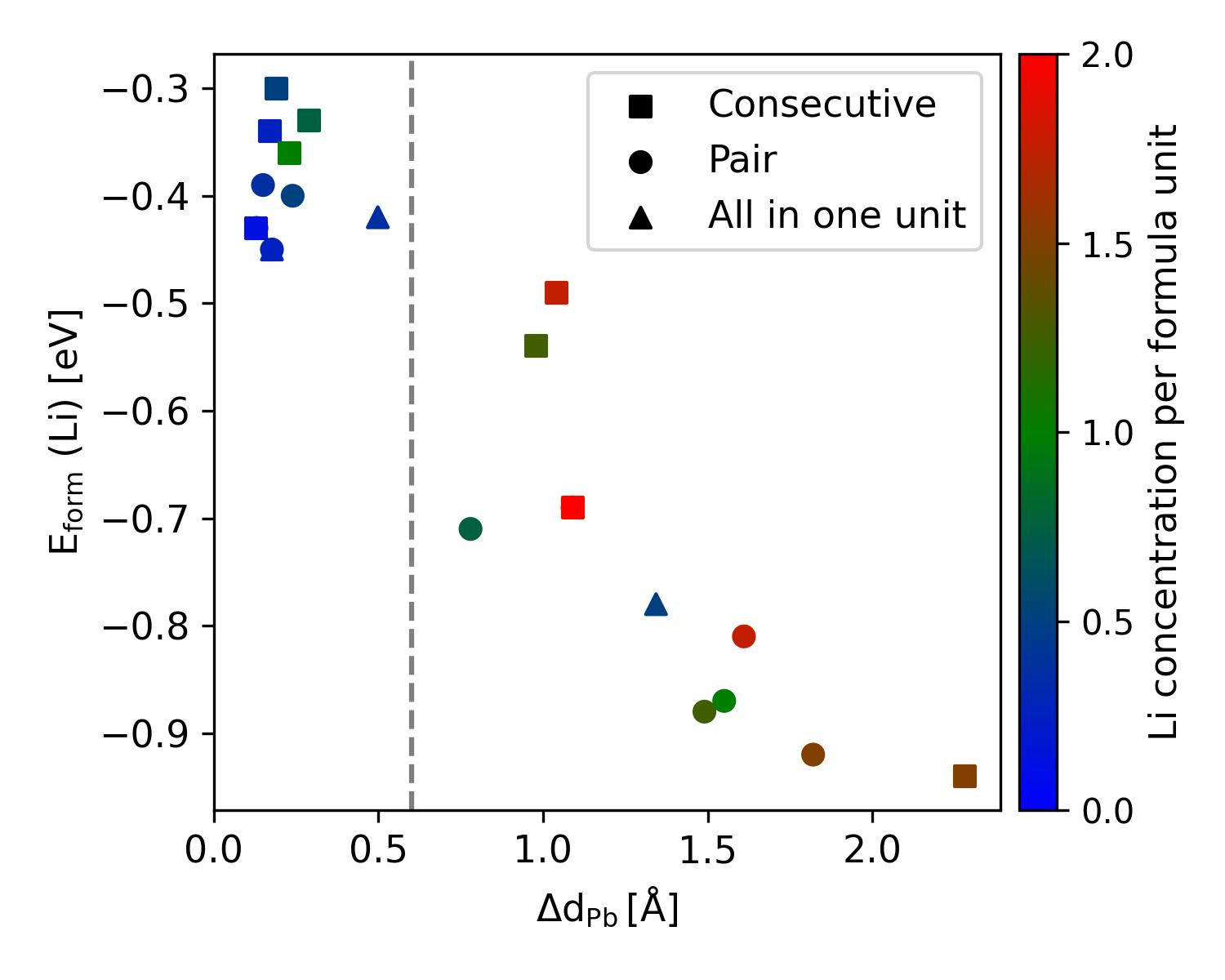}
	\caption{ The relationship between the formation energy of Li ions at T sites of the $\gamma$’ structure and the concomitant distortion of the pseudo-cubic perovskite structure, for different Li concentrations up to $x$=2. The distortion is quantified by the average displacement of Pb atoms at B sites, and the Li concentration is illustrated by the color code. Formation energies for Li ions at O sites are not included in this figure because their values are considerably higher. The symbols denote consecutive ($\square$), pairwise ($\bigcirc$) and all in one unit ($\Delta$) distributions of Li atoms, respectively.
	The vertical dashed line at d=0.6~\AA$ $ distinguishes perovskite-type crystal structures for d$<$0.6~\AA, which are weakly distorted by the inserted Li, from strongly distorted, no more perovskite-type crystal structures for d$>$0.6~\AA.
	 }
	\label{fig5}
\end{figure}

To further explore the potential concentration limit for the intercalation of Li into the $\gamma$’ structure, we investigate the Li concentrations up to $x$=2 at the T sites in the $\gamma$’ structure. In this structure, three ways to insert and arrange the Li ions are followed. The different arrangements of Li ions may have different formation energies in the perovskite crystal. The crystal model is a 2$\times$2$\times$2 supercell, comprising eight pseudo-cubic perovskite unit cells. Each cell is represented as a cubic cell marked by eight corner Pb atoms. Accordingly, these are the three arrangement ways: i) the "consecutive" arrangement means to add sequentially one atom to each cell until all eight cells contain one Li atom. And then, the procedure is repeated to add a second atom to each cell until all cells contain two Li atoms. ii) the "pair" arrangement stands for adding two Li atoms to the same cell at a time, followed by adding two atoms to another cell. This procedure is continued until all eight cells contain two Li atoms. iii) the "all in one unit" arrangement means to select a single cell and to add all Li atoms to that cell. In this arrangement, all Li atoms are concentrated in one cell, while the other seven cells remain devoid of Li atoms. Please note that our investigation here focuses on the potential concentration limit for Li in this perovskite crystal. It does not involve the methods of the process of inserting Li atoms or seek any controlled process to get a certain arrangement. The O sites are not considered in these arrangements because their formation energy is substantially higher, compared to that of the T sites. These strategies explore the stability of interstitial Li atoms from the most uniform to the most clustered Li distributions, which corresponds to the possibility of Li clustering in experiments.

Figure~\ref{fig5} illustrates the variation of the formation energy for different arrangements of Li with concentrations up to $x$=2. We observe a correlation between the formation energy and another type of lattice distortion. This lattice distortion can be quantified by the average displacement of Pb atoms from the B sites, denoted as $\Delta \rm {d_{Pb}}$. The pseudo-cubic lattice of the perovskite structure is formed by Pb atoms located at the B sites, thus this average displacement serves as an indicator of the degree of distortion from the pseudo-cubic perovskite structure. In that structure, a chemical bond distortion may become clearly recognizable when the lattice length changes by more than 10\%. Given that the lattice constant of the cubic perovskite structure is 6.27~\AA, we set 0.6~\AA~as the threshold for $\Delta \rm {d_{Pb}}$ to indicate a significant distortion away from the pseudo-cubic structure. When this threshold is exceeded, the pseudo-cubic framework undergoes such a deformation that the structure can no longer be classified as perovskite-like. Although some high Li concentrations are energetically favorable, a detailed study of such non-perovskite-type structures would deviate from the focus of our research. In the figure, we have marked this threshold with a vertical gray dashed line. Within the range that does not exceed this threshold, the concentration limit for Li uptake is found to be $x$=1. Any case with $x$$>$1 will have a severely deformed crystal structure. Hence, we infer that the concentration limit for the uptake of Li in the $\gamma$' structure of CsPbI$_3$ is $x$=1.

\subsection{Electronic structure analysis}

In the following we investigate the effect of interstitial Li atoms on the electronic band gap of the CsPbI$_3$ perovskite. We examine several factors, such as the Li-induced structural distortion, the electronic screening, and the defect states in the band gap  related to the interstitial Li atoms. We limit the examined Li concentrations to $x\leq 1$ due to the instability of the perovskite structure for higher Li concentrations. 

\begin{figure}
	\centering
	\begin{overpic}[width=0.45\textwidth]{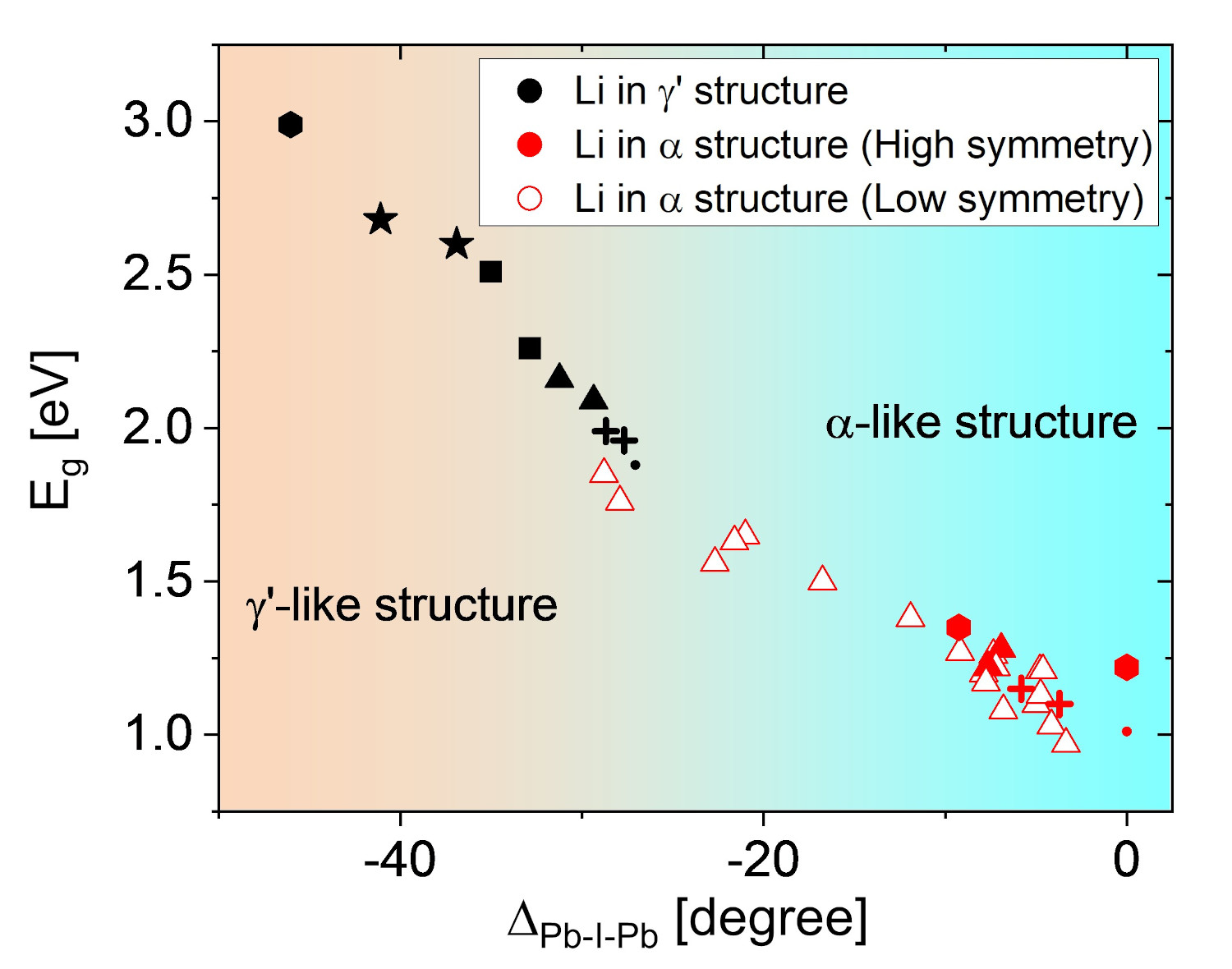}
		\put(2,70){(a)}
	\end{overpic}
	\begin{overpic}[width=0.44\textwidth]{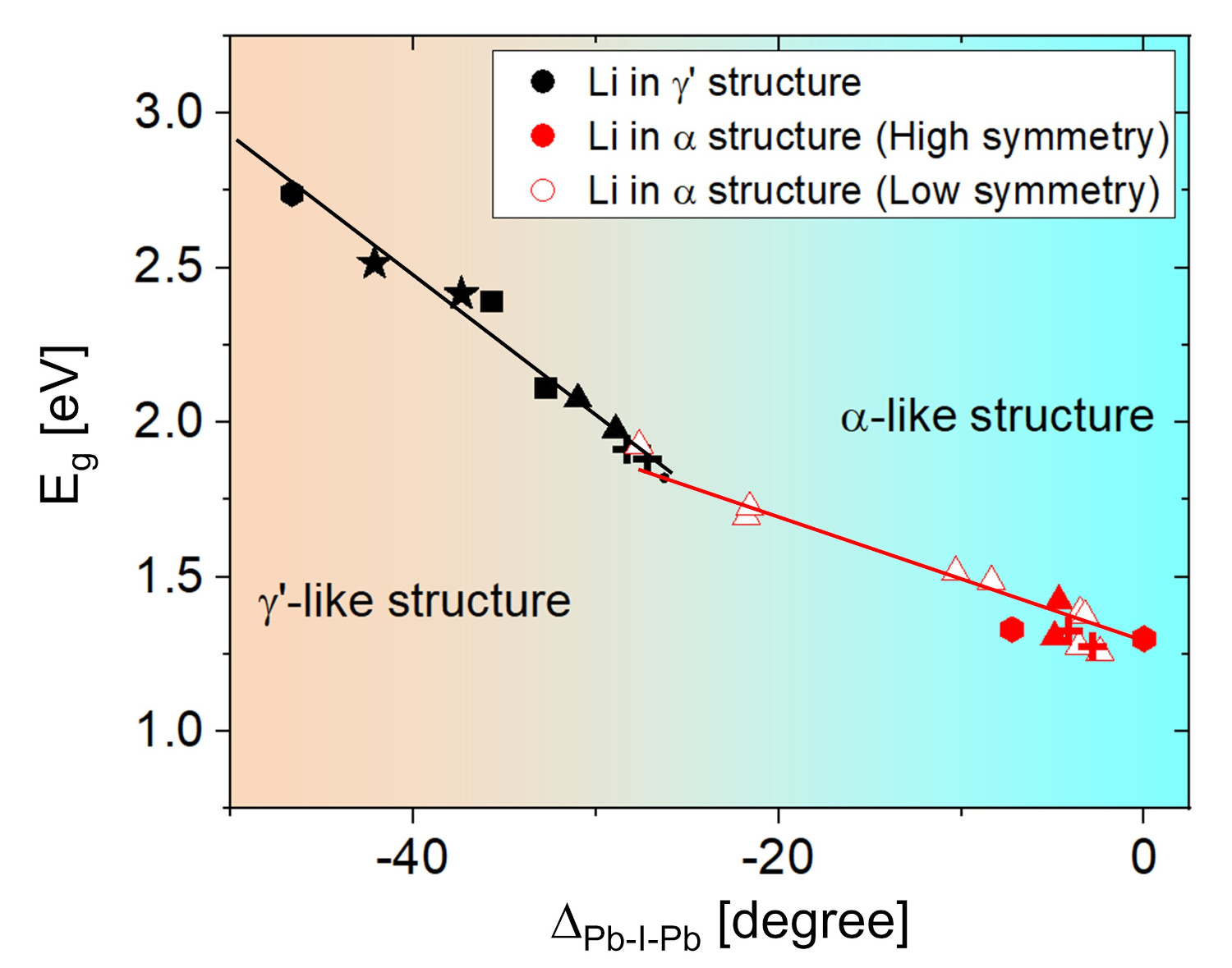}
		\put(2,70){(b)}
	\end{overpic}
	\caption{(a) Band gap E${\rm _g}$ vs. $\Delta_{\text{Pb}\mbox{-}\text{I}\mbox{-}\text{Pb}}$ for Li$_x$CsPbI$_3$ with different Li content $x$ and Li distribution.  Black and red symbols denote Li in the $\gamma$’ and $\alpha$ structures, respectively. Orange and light blue colored areas indicate the structural regimes, respectively. Different concentrations are indicated as $x$=0 (•), $x$=1/8 (+), $x$=1/4 ($\triangle$), $x$=1/2 ($\square$), $x$=3/4 ($\largestar$), $x$=1 ($\hexago$). Filled symbols refer to structures that retain $D_{4h}$ and $C_{3v}$ symmetry for O and T sites in the $\alpha$ structure, respectively. Other lower symmetry arrangements of Li are indicated by open symbols. (b) Band gaps obtained after subtracting the influence of $x$ extra electrons in the conduction band, estimated as $x$~$\cdot$~0.22~eV (see text for details). The black and red lines are linear fitting to corresponding data points.}
		\label{fig6}
\end{figure}

The relationship between the band gap E${\rm _g}$ and the variation of the bond angle $\Delta_{\text{Pb}\mbox{-}\text{I}\mbox{-}\text{Pb}}$ for the $\alpha$ and $\gamma$' structures is illustrated in Figure~\ref{fig6}a. We find a relationship between the two variables that appears linear across the high and low symmetry arrangements of Li in structures that were derived by adding Li into the $\alpha$ structure (red symbols). The structure with $x$=1 for the insertion of Li at the O site is an outlier, as it maintains the host crystal structure due to an artificially high symmetry. The band gap increases as the Li concentration rises. This growth trend is consistent with the earlier analysis of the structural distortion that arises when the Li content increases, i.e., increasing $x$ leads to an increase of $\Delta_{\text{Pb}\mbox{-}\text{I}\mbox{-}\text{Pb}}$, which in turn increases E${\rm _g}$.

In the $\gamma$’ structure, 
we observe a similar linear relationship between E$_g$ and $\Delta_{\text{Pb}\mbox{-}\text{I}\mbox{-}\text{Pb}}$ (black symbols).
However, the two linear regimes differ in slope and axis intercept.

The strong dependence of E${\rm _g}$ on $\Delta_{\text{Pb}\mbox{-}\text{I}\mbox{-}\text{Pb}}$, which is in line with the different band gaps of the $\alpha$ structure (1.29 eV, $\Delta_{\text{Pb}\mbox{-}\text{I}\mbox{-}\text{Pb}}$=0$^\circ$) and the $\gamma$' structure (1.82 eV, $\Delta_{\text{Pb}\mbox{-}\text{I}\mbox{-}\text{Pb}}$=-26$^\circ$), indicates that the structural changes dominate the influence that interstitial Li atoms have on the electronic structure of CsPbI$_3$. This is in line with the notorious sensitivity of band gaps in group-IV halide perovskite structures to the octahedral tilt angles \cite{knutson2005tuning}.

\begin{figure}
	\begin{overpic}[width=0.5\textwidth]{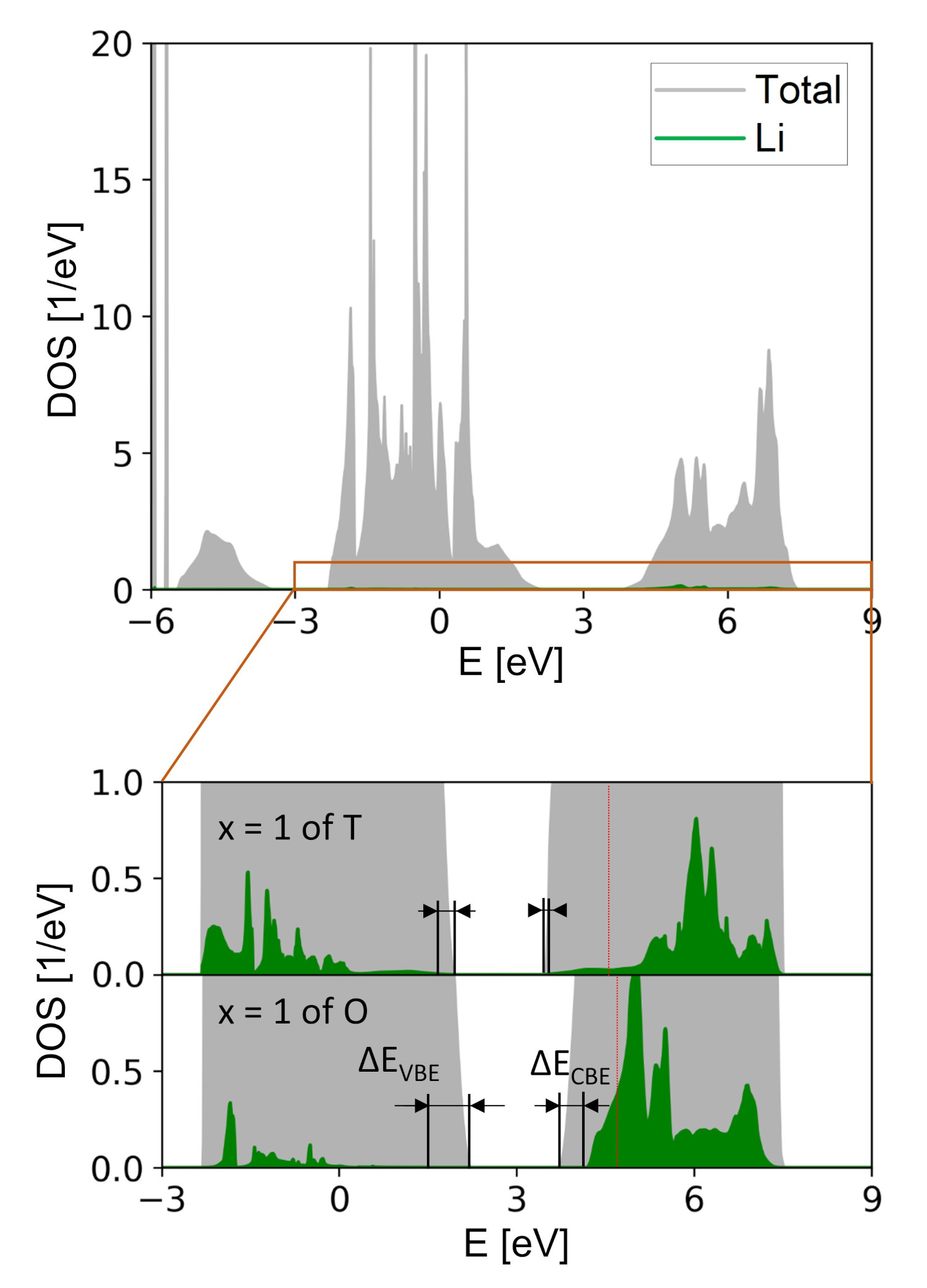}
		\put(2,95){(a)}
		\put(2,45){(b)}
	\end{overpic}
	\begin{overpic}[width=0.5\textwidth]{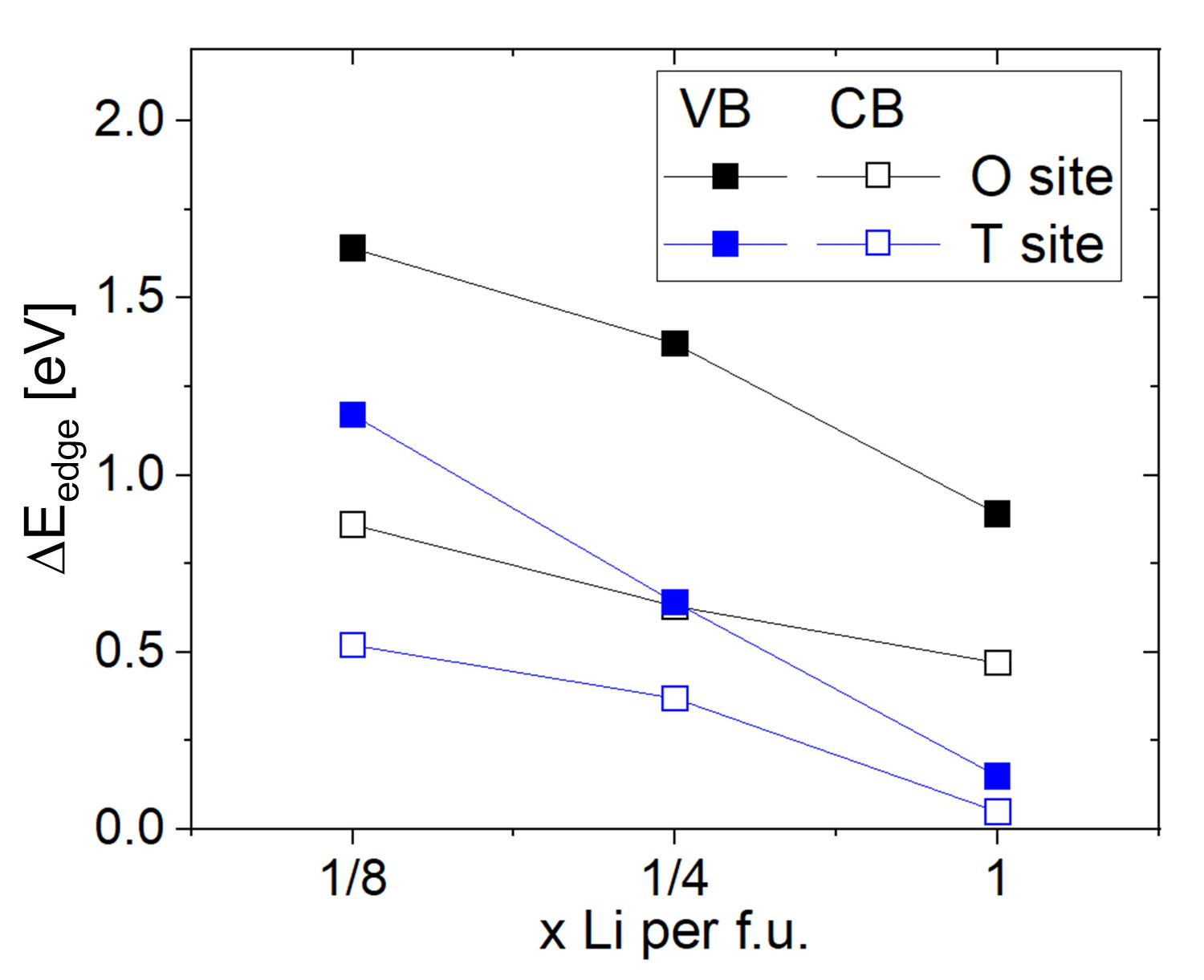}
		\put(-2,75){(c)}
	\end{overpic}
	\caption{(a) DOS of Li$_x$CsPbI$_3$ in the $\alpha$ structure and (b) the detailed DOS for interstitial Li on the O and T site with $x$=1. Structures with $D_{4h}$ and $C_{3v}$ symmetry are considered for O and T sites, respectively. Total DOS (gray) and Li projected DOS (green) are shown. $E_F$ is indicated by red dashed lines and the energy separation of bands containing Li contributions from the two band edges, $\Delta$$E_{\text{VBE}}$ and $\Delta$$E_{\text{CBE}}$, are marked by black lines and labels (c) The energy separation of bands with Li content from either band edge, $\Delta E_{\text{edge}}$ (edge: VBE or CBE), for Li$_x$CsPbI$_3$ with varying $x$ in the $\alpha$ structure.}
	\label{fig7}	
\end{figure}

To understand the direct effects of the insertion of Li atoms on the electronic structure of a CsPbI$_3$ crystal, we analyze the density of states (DOS) in the proximity of the band gap in Figure~\ref{fig7}a. Upon the addition of Li to the $\alpha$ structure, the Fermi level (E$_F$) shifts to the edge of the conduction band
for $x>0$. 
An interstitial Li atom adds its $1s$ orbital as a deep core state (not shown). 
The Li 2s orbital remains mostly unoccupied, but hybridizes with Pb and I orbitals in the regions of the conduction and valence band edges (CBE and VBE), respectively. For small concentration $x$ of Li, there is a significant distance of states with Li contributions to the two band edges ($\Delta E_{\text{VBE}}$ and $\Delta E_{\text{CBE}}$, cf. Fig.~\ref{fig7}b). With increasing $x$ of Li, this distance shrinks, cf. Fig.~\ref{fig7}c. 
This implies that Li affects the local environment of more I and Pb atoms as the value of x increases. When Li atoms are located on the T sites, their contribution to the DOS has a broader distribution in energy, compared to Li atoms at the O sites at the same concentration. This indicates a stronger hybridization of Li with the Pb and I orbitals on the T sites than on the O sites. 

We have also analyzed the DOS for various concentrations of interstitial Li in the $\gamma$' structure and obtain qualitative similarities to the findings for Li in the $\alpha$ structure. With a higher Li concentration, stronger structural distortions occur, resulting in a higher E${\rm _g}$. The Li $2s$ orbitals contribute to both the VBE and CBE at any concentration due to the stronger structural distortion, but their impact is still minor in magnitude.

\begin{figure}
	\centering
	\begin{overpic}[width=0.4\textwidth]{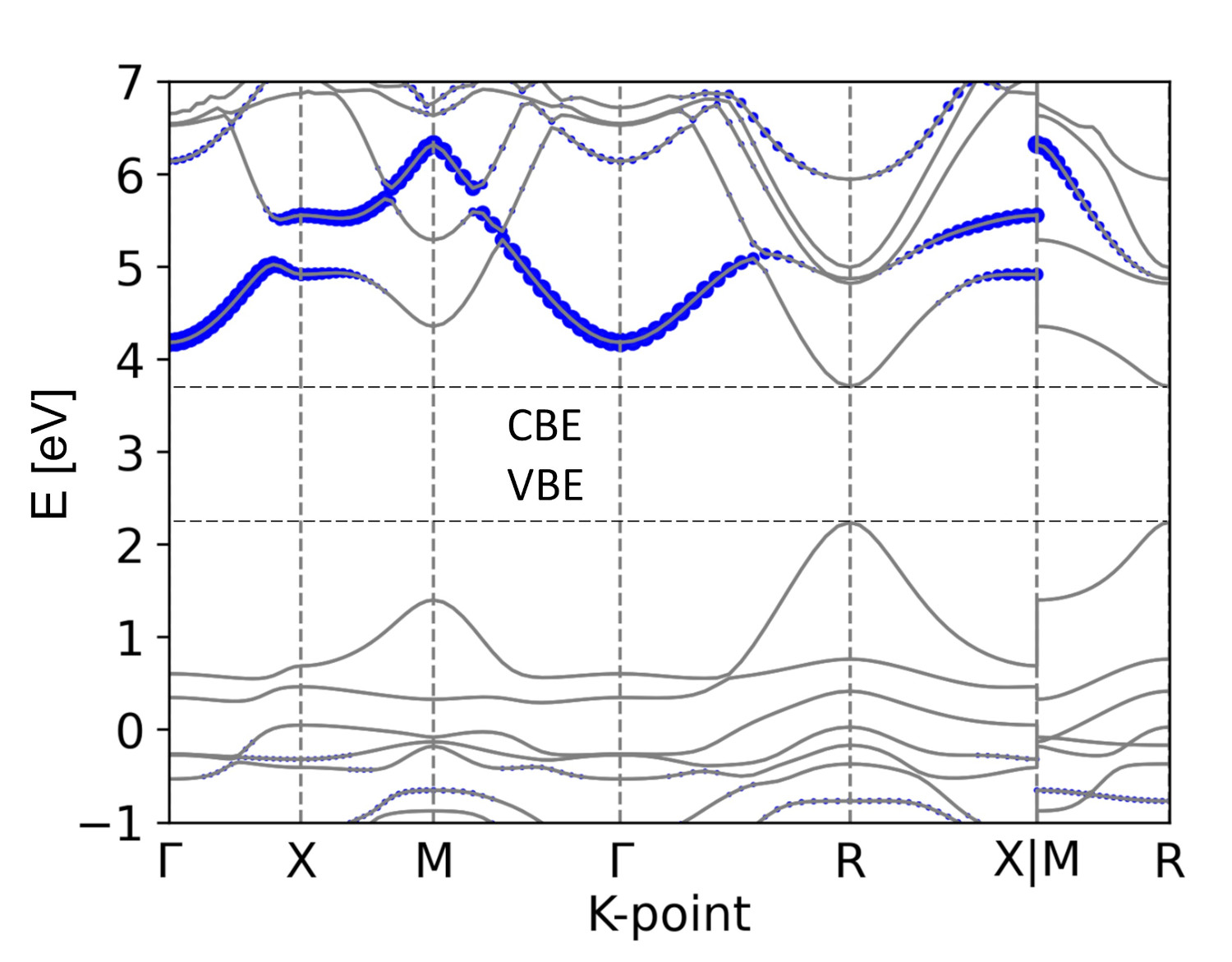}
		\put(-2,63){(a)}
	\end{overpic}
	\begin{overpic}[width=0.4\textwidth]{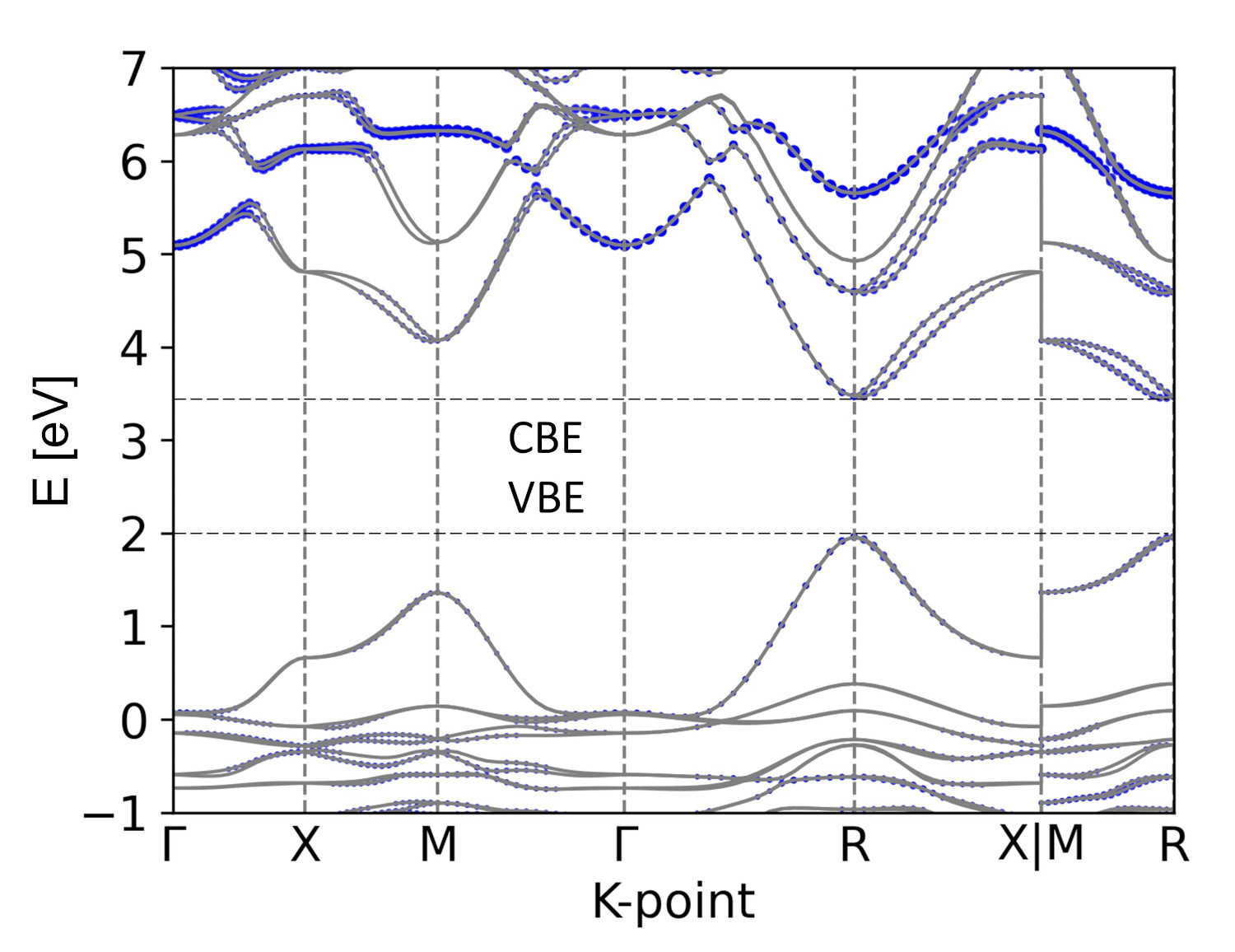}
		\put(-2,63){(b)}
	\end{overpic}
	\caption{Band structure of Li$_x$CsPbI$_3$ with Li atoms on the a) O and b) T sites for $x=1$ in the $\alpha$ structure. Li contributions to bands are indicated by blue circles. The circle radius indicates the magnitude of the contribution.}	\label{fig8}	
\end{figure}

Figure~\ref{fig8} displays the band structures of $\alpha$-Li$_x$CsPbI$_3$ with Li content $x$=1 at O and T sites to further analyze the hybridization between Li 2s orbitals and the host-crystal orbitals. The Li contribution to the bands is indicated by blue circles. When the Li atoms are located on O sites, the Li $2s$ states form a recognizable defect level (the bands with dense blue circles) that is more localized and has no contribution to the band edges. On the T site, the hybridization between Li $2s$ orbitals and Pb $6p$ orbitals occurs over a broader energy range. 

The strength of the orbital overlap can be estimated based on the distances from Li to I and Pb. The nearest Li--I and Li--Pb distances of Li on T sites are  2.61--2.66~\AA~ and 2.79--3.03~\AA, respectively. For Li on O sites, the Li--I and Li--Pb distances are larger, 2.85--3.11~\AA~and 4.40--4.47~\AA, respectively. As a result, the increased overlap estimated from the smaller spatial distance in case of the T site is consistent with a broadened Li $2s$ contribution that extends closer to the band edges for Li on T sites. However, in either case, the Li $2s$ contribution to the band edges is negligible compared to the manifold of I and Pb states, cf. Fig.~\ref{fig7}a.  

\begin{figure}
	\includegraphics[width=0.42\textwidth]{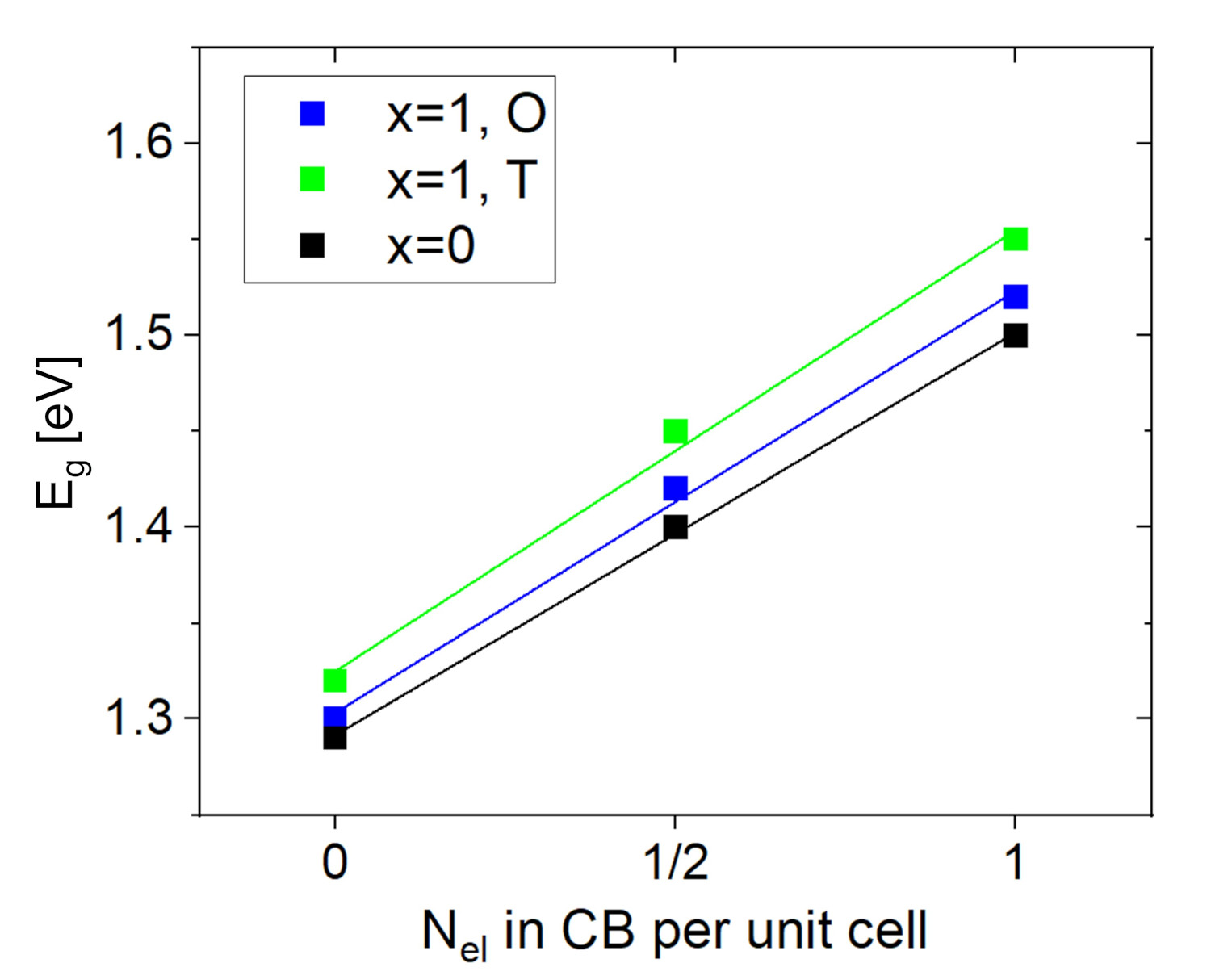}
	\caption{Band gap change with varying number of electrons in the conduction band for three $\alpha$ structures of Li$_x$CsPbI$_3$ with $x=$ 0 and $x=$ 1 with Li atoms at O or T sites. The straight lines are linear fitting to the corresponding data points.}
	\label{fig9}	
\end{figure}

So far, we have analyzed the case of inserting neutral Li atoms into the neutral perovskite crystal. Besides the $2s$ orbital, which appears to have a weak effect on the band edges, the insertion of a Li atom also introduces an additional electron into the CsPbI$_3$ crystal. In this scenario, the Fermi level (E$_F$) shifts into the conduction band, and the band gap of CsPbI$_3$ with Li in the O sites changes by 0.22 eV from $x=0$ to $x=1$, even without any structural alterations. This indicates that the extra electron in the conduction band directly influences the electronic structure. We quantify this effect in three different $\alpha$ structures, the one for $x=0$ as well as the cases $x=1$ with Li in O and T sites, respectively. Here, we keep the crystal structures unchanged when varying $N_{\rm el}$.

The relationship between E$_g$ and the number of excess electrons per perovskite formula unit, $N_{\rm el}$, is illustrated in Figure~\ref{fig9}. In all three investigated cases, the band gap increases with $N_{\rm el}$; this behavior is rather well described by a linear relation. The three slopes vary only between 0.21 and 0.23 eV per electron. This means that the influence of extra electrons at the CBE in Li$_x$CsPbI$_3$ alloys on the band gap can be modeled by an average shift of approximately 0.22 eV per extra electron.

We can use this relationship to subtract the influence of the extra charges by subtracting $x \cdot 0.22$~eV from the computed band gap values. Doing so, our computed structures serve as representative model structures for dilute Li$_x$CsPbI$_3$ alloys with very small $x$, where the structural changes of Li dopants occur locally, but the extra electrons may be delocalized. This description of this dilute limit is, however, not our main focus, having a battery application and, thus, large $x$ in mind. 

In the concentration range of interest, we can use the above relation to isolate the structural influence on the band gap from the two influences of Li (extra electrons and band hybridization) that also contribute to the relation between $E_g$ and $\Delta_{\text{Pb}\mbox{-}\text{I}\mbox{-}\text{Pb}}$. 

The resulting behavior without the additional influence of extra electrons is displayed in Figure~\ref{fig6}b. 
As noted above, the region of $\gamma'$-like structures is well represented by a linear relationship. Likewise, the structures with Li being added to the $\alpha$ structure are described well by a linear relationship. Accounting for the influence of a varying amount of extra electrons the outlier mentioned above for Li on the O site with $x$=1 fits well now to the linear relation.

We also analyzed a removal of all Li contributions, i.e., we determined the CsPbI$_3$ structures with the distortions caused by Li atoms, and then we removed the Li atoms when evaluating the electronic structure (i.e., the extra electrons, the $2s$ bands, and the ionic cores of Li are omitted). We find that the influence on $E_g$, coming only from the structural changes of the CsPbI$_3$ lattice in that way does only in some cases correlate well with the band-gap data given in Figure~\ref{fig6}b. In about half of the cases, the interactions that are omitted when removing Li as bonding partner lead to significant changes in the local DOS of the Iodine $p$ orbitals that are directed towards Li. Thus, the change of $E_g$ with the Li content is dominated by the structural changes of the CsPbI$_3$ lattice, cf. Fig.~\ref{fig6}a, and one can describe these changes by separating the structural effect from the varying amount of extra electrons in the conduction band, cf. Fig.~\ref{fig6}b. But describing the observed band-gap variation purely by the structural changes of the CsPbI$_3$ lattice and neglecting the presence of Li as bonding partner is not successful.

\section{Discussion} \label{discussion}

The ordered cubic $\alpha$ structure has served as the structural model of choice in most previous theoretical studies \cite{zhang2019improved,kye2019vacancy,jong2018first}. However, recent studies \cite{gebhardt2021efficient,wiktor2017predictive,yang2017spontaneous,whitfield2016structures} suggest that this model is probably not the preferred structural representation of the dynamic CsPbI$_3$ crystal around room temperature and the $\gamma’$ structure is a more reasonable model.  We find that the latter choice is even more relevant when adding Li to the structure, which itself promotes the occurrence of octahedral tilts in our results, cf. Fig.~\ref{fig3}. 

According to our simulation results, the limit for Li uptake into a hypothetical high symmetry $\alpha$ structure would be $x$=1/4, cf.~Fig.~\ref{fig2}. This is in line with the DFT results of Dawson et al.~(from $x$=0.037 to $x$=1) on MAPbI$_3$ and MAPbBr$_3$  \cite{dawson2017mechanisms}. These results indicate that both hybrid halide perovskites cannot bear a Li concentration of $x$=1, and perovskite dissolution or distortion is energetically preferred. Similarly, Büttner et al.~\cite{buttner2022halide} reported that the electrolytes containing Li dissolve the perovskite immediately, and they concluded the impossibility to store Li ions in the hybrid halide perovskites. This can be explained with the ability of the organic cations in hybrid perovskites to form hydrogen bonds to polar solvent molecules, which facilitates the perovskite decomposition \cite{zheng2019unraveling}. This decomposition is even thermodynamically favorable for Pb-based hybrid halide perovskites \cite{zhou2019chemical}. 
In contrast, the inorganic CsPbI$_3$ perovskites are  thermodynamically stable \cite{zhou2019chemical}.

As opposed to the case of the $\alpha$ structure, the formation energy of interstitial Li in the $\gamma$' structure increases with increasing $\Delta_{\text{Pb}\mbox{-}\text{I}\mbox{-}\text{Pb}}$ for the most stable O and T sites  at low Li concentration, cf.~Fig~\ref{fig4}. Thereafter, for $x$=1/2, the formation energy reaches a plateau, while the structural distortion still increases with increasing concentration. Overall, the energy profile is in line with a stabilization of the $\gamma$' structure, with the lowest energy obtained for the Li-containing structure that is closest to the $\gamma$' structure. The formation energies in a range from -0.43~eV to -0.30~eV (from $x$=1/8 to 1) indicate the potential for battery applications.
 
Higher Li concentrations with $x>1$ disrupt the perovskite structure, altering the cubic lattice by the displacement of Pb ($\Delta{\rm d_{Pb}}$) instead of only changing the Pb-I-Pb angles ($\Delta_{\text{Pb}\mbox{-}\text{I}\mbox{-}\text{Pb}}$), cf. Fig~\ref{fig5}. This structural distortion differs significantly from the conditions of Li in the $\alpha$ structure and the lower Li concentrations in the $\gamma$' structure. We propose that it is unfeasible to maintain the perovskite structure in high concentrations of Li, whereby the limiting concentration of Li is $x$=1.

Interestingly, as the concentration of Li increases beyond $x$=1, the structural change is accompanied by a lower formation energy than for $x\less$ 1. It indicates a high capacity of the material for Li storage. Similar results have been reported from experiments \cite{xia2015hydrothermal,vicente2017methylammonium}. Xia et al. \cite{xia2015hydrothermal} demonstrated an astonishingly high Li concentration with $x\approx6$ in Li$_x$MAPbBr$_3$, although the Li capacity faded rapidly after only 30 charge/discharge cycles. Vicente et al. \cite{vicente2017methylammonium} reported that the Li concentration in Li$_x$MAPbBr$_3$ reaches a value as high as $x=3.7$. The rapid fading of the high Li capacity indicates structural distortion away from the perovskite structure, which is in line with our results for high Li concentrations.

Finally, we found that the insertion of Li increases the band gap of the CsPbI$_3$ perovskite, cf.~Fig.~\ref{fig6}. Similar results for Li containing perovskite compounds were reported from photoluminescence experiments \cite{wu2021dual,jiang2017electrochemical,mathieson2022solid}. Jiang et al.~\cite{jiang2017electrochemical} found a blue shift in the photoluminescence spectrum by Li inserted in CsPbBr$_3$, and attributed it to the Burstein–Moss effect \cite{burstein1954anomalous}. This effect refers to the phenomenon that electrons at the VBE require higher energies to be excited due to the occupation of the CBE, thus exhibiting a higher absorption band gap. A similar blue shift was also found in Li$_x$MAPbBr$_3$ by Mathieson et al.~\cite{mathieson2022solid}. They suggested that although the inserted Li atoms narrow the band gap, the electrons can only be excited by photons of higher energy due to the occupied CBE. However, the influence of the Li atoms on the band gap of the perovskite crystal is multifaceted, and the mechanism mentioned in the cited literature is only about the occupied CBE. According to our results, the main reason for the increased band gap is the inevitable structural distortion induced by the Li atoms.

The band gap increase caused by the inserted Li is determined by the covalent bonding character of the halide perovskite. The VBE and CBE of CsPbI$_3$ are mainly composed of the covalent $\sigma$*-bonding states of $p$ and $s$ orbitals of Pb and I \cite{umebayashi2003electronic}, and the dispersion of the bands is determined by the overlaps of these orbitals. The insertion of Li atoms increases the  $\Delta_{\text{Pb}\mbox{-}\text{I}\mbox{-}\text{Pb}}$ by attracting I atoms and, thus, reducing the overlap between the Pb and I orbitals. 
As a result, the presence of Li significantly increases the band gap, finally rendering three-dimensional halide-perovskite crystals not suitable for the use in photo-battery devices. But the organic-inorganic hybrid layers in two-dimensional perovskites may be more promising for Li storage and photovoltaic function. Ahmad et al. demonstrated the insertion of Li ions into a perovskite compound with a capacity of approx. 100 mAh/g and without compromising the photovoltaic efficiency \cite{ahmad2018photo}. Li is stored in the organic layer of the two-dimensional perovskite, which does not affect the band gap of the inorganic perovskite layer. Furthermore, Chen et al.~\cite{chen2022organic} proposed a plausible Li storage mechanism involving carbonyl groups. The aforementioned mechanism (or related ones) holds the potential for developing an integrated halide-perovskite photo-battery.

\section{Summary\protect} \label{summary}

In this work, the structural stability of CsPbI$_3$ when Li is intercalated, and the effect that Li has on the electronic structure of the resulting Li$_{x}$CsPbI$_{3}$, are investigated. By exploring Li$_x$CsPbI$_3$ compounds with various concentrations and arrangements of Li ions, we find that structural distortions need to be considered for getting the Li formation energies to identify the Li concentration limit, and that the distortion of Pb-I-Pb bond angles is a more influential key feature for the band gap increase of Li$_x$CsPbI$_3$ materials than just the Li concentration.

We analyze the structural stability as function of the Li insertion by two structural models for CsPbI$_3$: first, the cubic perovskite $\alpha$ structure, which is a hypothetical structure to be understood as a reference model with negligible probability to occur at room temperature; second, a distorted $\gamma$' structure, closely related to the $\gamma$ phase, that represents the lowest energy structure at room temperature. 
The hypothetical $\alpha$ structure energetically disfavors Li uptake and is likely to be structurally unstable for $x$ $>$ 1/4. Thus, such a theoretical model is unsuitable for interpreting
experiments that demonstrate the existence of Li$_x$CsPbI$_3$ compounds.
In the $\gamma$’ structure, interstitial insertion of Li is energetically
favorable.
Concentrations up to $x=1$ are accessible while keeping the perovskite host structure intact. In principle, forming the compound with even higher Li concentration is feasible. However, our investigation of structures with up to $x=2$ indicate that in such cases significant distortions of the host lattice structure occur. The structural distortions of such phase transitions and their potential irreversibility require further investigations.

For the concentration range $0\leq x \leq 1$,  the interstitial Li has several effects on the electronic structure of Li$_x$CsPbI$_3$: i) the induced structural distortion reduces the band dispersion, leading to a significant increase of the band gap; ii) the electronic screening of extra electrons in the conduction band leads to a slight increase of the band gap; iii) although the Li $2s$ orbitals hybridize with I and Pb orbitals, their effect on the band gap and the band edges is negligible. 

Altogether, these effects result in an increased band gap in Li$_x$CsPbI$_3$ compared to CsPbI$_3$. The band gap E${\rm _g}$ increases linearly with the induced averaged angle distortion $\Delta_{\text{Pb}\mbox{-}\text{I}\mbox{-}\text{Pb}}$ in the two structural regimes of the $\alpha$ and  $\gamma'$ models. For $x$=1, this leads to E${\rm _g}$ of 2.96~eV. This continuously increasing band gap needs to be taken into account for potential solar cell applications.

\begin{acknowledgments}
This work was supported by the Deutsche Forschungsgemeinschaft (DFG, German Research Foundation) under Germany’s Excellence Strategy-EXC-2193/1-390951807 (LivMatS). We thank the State of Baden-Württemberg (Germany) through bwHPC for computational resources.
\end{acknowledgments}



\providecommand{\noopsort}[1]{}\providecommand{\singleletter}[1]{#1}%

\end{document}